\documentclass[10pt]{bmc_article}

\usepackage{cite} 
\usepackage{url}  
\usepackage{ifthen}  
\usepackage{multicol}   
\usepackage[utf8]{inputenc} 
\usepackage{graphicx}
\newcommand{\angstrom}{\text{Å}}

\usepackage{amsmath,amssymb}

\newboolean{publ}

\newenvironment{bmcformat}{\begin{raggedright}%
\baselineskip20pt\sloppy\setboolean{publ}{false}}{\end{raggedright}%
\baselineskip20pt\sloppy}%

\begin{document}
\begin{bmcformat}
\title{ClusCo: clustering and comparison of protein models}

\author{Michal Jamroz$^{1}$
       \email{Michal Jamroz - jamroz@chem.uw.edu.pl}%
      and
         Andrzej Kolinski\correspondingauthor$^{1}$%
         \email{Andrzej Kolinski\correspondingauthor - kolinski@chem.uw.edu.pl}
      }
\address{
    \iid(1)Laboratory of Theory of Biopolymers, Faculty of Chemistry, University
of Warsaw, Pasteura 1, 02-093 Warsaw, Poland
}%

\maketitle

\begin{abstract}
        \paragraph*{Background:}  The development, optimization and validation
of protein modeling methods  require efficient tools for structural comparison.
Frequently, a large number of models need to be compared with the target native
structure. The main reason for the development of Clusco software was to create
a  high-throughput tool for all-versus-all comparison, because calculating
similarity matrix is the one of the bottlenecks in the protein modeling
pipeline.  
      
        \paragraph*{Results:} Clusco is fast and easy-to-use software for
high-throughput comparison of protein models with different similarity measures
(cRMSD, dRMSD, GDT\_TS, TM-Score, MaxSub, Contact Map Overlap) and clustering of
the comparison results with standard methods: K-means Clustering or Hierarchical
Agglomerative Clustering. 

        \paragraph*{Conclusions:} The application was highly optimized and
written in C/C++, including the code for parallel execution on CPU and
GPU version of cRMSD, which resulted in a significant speedup over similar
clustering and scoring
computation programs.
\end{abstract}

\ifthenelse{\boolean{publ}}{\begin{multicols}{2}}{}

\section*{Background}

The development, optimization and validation of protein modeling methods require
efficient tools for structural comparison. Frequently, a large number of models
need to be compared with the target native structure. There are numerous
measures of model similarity. The most popular is the cRMSD -- coordinate Root
Mean-Square Deviation (after the best superimposition)\cite{kabsch}. The other
popular scores are: GDT\_TS -- Global Distance Test Total Score\cite{Zemla2003},
MaxSub -- Maximal Substructure\cite{Siew2000}, TM-Score -- Template-Modeling
Score\cite{Zhang2004a}, or dRMSD -- distance Root Mean-Square
Deviation\cite{Torda1994}.

One of the methodologies most widely used for protein modeling includes
performing the clustering step after generating a protein conformation
ensemble\cite{Shortle1998a,Kolinski2004b,Roy2010,Xu2012,Rohl2004} followed by
the selection of a representative model (or models) for refinement.   To achieve
this, we need a similarity matrix of the whole ensemble, which contains
all-versus-all comparison (for $N$ conformers it gives $N(N-1)/2$ of score
calculation). However, many available applications are not optimized for running
time, because they were developed rather for simple pair comparison.

The main
reason for the development of Clusco software was to create a 
high-throughput tool for all-versus-all comparison,  because calculating
similarity matrix is the one of the bottlenecks in the protein modeling
pipeline.

\section*{Implementation}

The implementation of the similarity measures was performed  using OpenMP API
which supports multiprocessing programming. Additionally, the cRMSD algorithm
was coded on the Graphic Processor Unit (GPU) architecture using
\textsc{nvidia cuda}, which gave an over ten fold speedup in comparison with one
CPU.

We used an open source parallel K-means\cite{kmeans} clustering code,
implemented with OpenMP and serial C Clustering Library\cite{DeHoon2004} for
Hierarchical Agglomerative Clustering (single-linkage, maximum-linkage,
average-linkage).

\subsection*{Algorithms}
{\bf cRMSD} Coordinates Root Mean-Square deviation is defined as:
\begin{equation}
cRMSD=\sqrt{\frac{1}{N} \sum_{i=1}^N \| \mathbf{x}_i^A - \mathbf{x}_i^B \|^2}
\end{equation}

after the optimal superimposition. In this equation $N$ is the number of atoms, 
$x_i^A$ - $i$-th atom position  vector of protein $A$. There is no need to
superimpose structures (calculating rotation matrix) to obtain the cRMSD. By the
diagonalization of the $3\times 3$ covariance matrix $M$, we obtain the cRMSD
value by\cite{Bruschweiler2003}:

\begin{equation}
 \mathrm{cRMSD} = \sqrt {R_A^2 + R_B^2 - 2(\sqrt{\lambda_1}+\sqrt{\lambda_2}+S\sqrt{\lambda_3})}
\end{equation}
where $R_A$ is the radius of gyration of protein $A$, $S$ is the sign of the
covariance matrix  determinant, $\lambda$ is the eigenvalue (sorted in
descending order) of the square of the covariance matrix. These eigenvalues can
be computed by finding roots of the cubic equation instead of computational
demanding Singular Value Decomposition of the covariance matrix.

{\bf GDT\_TS} Global Distance Test Total Score is defined as:
 \begin{equation}
 \begin{split}
  \mathrm{GDT\_TS} =& \frac{1}{4}\left(\max C_{1\angstrom} + \max C_{2\angstrom} + \max C_{4\angstrom} \right. \\
    & \left. + \max C_{8\angstrom}\right)
 \end{split}
 \end{equation}

where $C_{1\angstrom}$ is the number of atom pairs below the $1\angstrom$
distance.  Max denotes here the maximal value for a series of superimpositions. 

The Global Distance Test algorithm is NP-hard\cite{Li2011a}, and all the GDT\_TS
computing algorithms use their own heuristics. Our GDT\_TS algorithm is as
follows: 1) divide the chain into all possible continuous $4, N/4, N/2, N$
fragments, and 2) use them as initial superimposition fragments (i.e.
superimpose the whole structure by a rotation matrix computed by superimposing
each fragment), 3) find atom pairs which are closer by cutoff ($1, 2, 4, 8
[\angstrom]$), 4) select atoms which are closer than $ 3.5 \angstrom$ and use
them for another superimposition until the number of selected atoms does not
change in four iterations.

{\bf TM-Score, MaxSub} (Template Modeling Score and Maximum Subset,
respectively) \cite{Zhang2004a, Siew2000} -- both scores are variations of
Levitt and Gerstein (1998) score.

TM-Score is defined as:
\begin{equation}
 \mathrm{TM-Score} = \max\left[ \frac{1}{N} \sum_{i}^N \frac{1}{1+\left(\frac{d_i}{d_0}\right)^2} \right]
\end{equation}

where $d_0 = 1.24\sqrt[3]{N - 15} - 1.8 [\angstrom]$, and MaxSub is defined as
the TM-Score  with $d_0 = 3.5\angstrom$. For the calculation of both scores we
used the same searching algorithm as for GDT\_TS, which means that the
computational costs of GDT\_TS, MaxSub, and the TM-Score will be the same.

{\bf dRMSD} distance Root Mean-Square Deviation\cite{Torda1994}. This score is
the  deviation of inter-molecular distance matrices:
\begin{equation}
 \mathrm{dRMSD} = \sqrt{\frac{2}{N(N-1)}\sum_{i=1}^{N-1}\sum_{j=i+1}^{N}\left(d_{ij}^A - d_{ij}^B\right)^2}
\end{equation}

where $d_{ij}^A$ is the distance between $i$ and $j$ atoms of protein model $A$.
 Note that the representation of molecule as a distance matrix causes loss of
information about chirality.

{\bf CMO} Contact Map Overlap\cite{Fraser2007}. Using a representation of the 
protein structure as a binary matrix $C$, defined as: 
\begin{equation}
C_{i,j} = \left\{ \begin{array}{ll}
1 & \mbox{if } | \mathbf{x}_i - \mathbf{x}_j| <\mbox{cut-off}\\
0 & \mbox{otherwise}
\end{array}\right.
\end{equation}
we use Sørensen Similarity Index as a similarity score between the two proteins $A$, $B$:
\begin{equation}
 S = 100\frac{2 n(A \cap B)}{n(A)+n(B)}
\end{equation}
where $n(A)$ is the number of contacts in protein $A$.

\section*{Results and Discussion}

Using the cRMSD, dRMSD, MaxSub, GDT\_TS, TM-Score, CMO as a similarity measure, 
Clusco can calculate all-versus-all (or with respect to the reference model)
scores of proteins from a one-column list file or using multimodel pdb file. The
calculated results may be used, for example, as a similarity matrix input for
clustering algorithms or clustered by Clusco itself.

In this section we show the examples of usage and the performance of Clusco with
respect to other similar programs. All tests were performed on a box with
\textsc{intel} E5649 CPU (24 threads), \textsc{nvidia} GeForce GTX 470 GPU and
24GB of RAM. The computation time elapsed was assessed by the standard *NIX
``time'' program.

\subsection*{Selecting of pairs of models within a given cRMSD threshold}

Recently, Fogolari and coworkers\cite{Fogolari2012} described an algorithm for 
reducing the computational cost of all-versus-all comparison of protein models
using cRMSD by inverse triangle inequality. As an example of that idea, the
authors provided \textsc{fsss} software and 1ctf protein models from
4state\_reduced decoy set (Decoys 'R' Us)\cite{Samudrala2000} as an input
ensemble. The \textsc{fsss} software outputs pairs of models with cRMSD below a
given threshold (3.2$\angstrom$ in this example).

We compared \textsc{fsss} and Clusco based on this dataset, recording the time 
spent by one CPU performing the task. Note that Clusco computes all-versus-all
scores by default, and to get results similar to the ones obtained from
\textsc{fsss} (only pairs below given threshold) we needed to filter the output
by awk (standard *NIX program):
\texttt{clusco -l 4state\_reduced\_1ctf.list -s rmsd -o 4state.tmp;} 
\texttt{awk '{\char '173}if(\$3<3.2) print \$line{\char '175}'\  4state.tmp > output.rmsd}

\textsc{fsss} software spent 149.14 seconds on this task, and Clusco+awk spent 
0.46~(Clusco) + 0.11~(awk) seconds, which amounts to  $261\times$ speedup. Note
that it is possible to improve these results, using parallel execution of Clusco
(by simply defining shell variable \texttt{OMP\_NUM\_THREADS} before Clusco
execution).

\subsection*{Clustering of protein decoys from five independent Molecular
Dynamics trajectories}

The decoy set vhp\_mcmd\cite{Fogolari2005} from Decoys 'R' Us database contains
the  results of five (NATIVE, F1, F3, F4, F7) Molecular Dynamics simulations of
the thermostable subdomain from chicken villin headpiece (36 residues, pdb code:
1vii), starting from different protein conformations. The set contains 6256 of
villin conformations in total, in the range of 0.49 - 12.8 $\angstrom$ cRMSD to
the experimental structure deposited in the Protein DataBank.

Using cRMSD and each of the Clusco clustering schemes, it is possible to
separate  this decoy set roughly into former trajectories, as we show in SI
Table 1. Each of the Hierarchical Agglomerative methods divides decoys into
rather separate clusters i.e. more than 85\% of trajectory models create a
separated cluster, while of ``NATIVE'' and ``F1'' models create one common
cluster (which is the result of ``F1'' convergence to the native structure
during simulation -- we refer the interested readers to Figure 2 in
\cite{Fogolari2005}). Other clustering scheme in K-means results in the grouping
of ``NATIVE'' and ``F1'' models into four separated clusters.

Command to perform clustering described above: 
\texttt{clusco -l vhp\_mcmd.list -s rmsd <0,1,2,3> 8}

The running time for this dataset varied from 5 seconds (for K-means clustering 
and GPU cRMSD comparison), to 3.5 minutes (for average-linkage, Hierarchical
Agglomerative Clustering and CPU cRMSD comparison,  see Additional file 1 Table
1.

\subsection*{Selecting representative model from {\it de-novo} modeling decoy set}

Clustering of protein models after \textit{de-novo} simulations is one the
methods most commonly used for the selection of the representative model from
the decoys set\cite{Shortle1998a,Kolinski2004b,Roy2010,Xu2012,Rohl2004}. We
compared Clusco with other clustering software
(\textsc{durandal}\cite{Berenger2012}, \textsc{calibur}\cite{Li2010},
\textsc{spicker}\cite{Zhang2004b}) in terms of results and computation time. To
do this, we used public available I-Tasser\cite{Wu2007c} decoys set, containing
12500-32000 models for each of 56 modelled target protein.

\textsc{calibur} uses preprocessing of decoy set in three ways:
screening-out unlikely candidates by setting lower and upper cRMSD bounds, using
triangular inequality for assessing if particular model is within the threshold
distance from a group of models (which reduces the number of structure
comparisons), detecting and ignoring outlier decoys. \textsc{durandal} uses
triangular inequality (like \textsc{calibur}) for the approximation of cRMSD
value of randomly chosen decoy and fill-up distance matrix until it contains
proper amount of information for the next, clustering step. \textsc{spicker}
selects the decoy with the largest number of structurally similar decoys (by
automatically detected threshold value) and creates the first cluster. The
process is repeated to get a sufficient number of clusters.

Clusco was run with cRMSD as a similarity measure (just as \textsc{durandal},
\textsc{calibur} and \textsc{spicker}) and K-means as a clustering method. We
set number of clusters to 20:
\texttt{clusco -l list -s rmsd 0 20}

The Clusco representative model was selected by $\min(\left<R\right>/f)$, where
$f$ denoted the fraction of elements in particular cluster and $\left<R\right>$
-- the average cRMSD between cluster elements. 

For the comparison of software reliability, we calculated tm-score to the
experimental structure (Additional file 1 Table 2) and Z-score of the tm-score
(where Z-score $< 0$ means that a model is worse than the average structure of
the decoy set, for detailed results see Additional file 1 Table 3).

According to the average tm-score, all of the programs gave similar results:
all, except \textsc{durandal}, gave the average tm-score 0.59, and
\textsc{durandal} gave the score of 0.58. According to Z-score, the best
algorithms were \textsc{calibur} and Clusco (49/56 of the models with Z-Score
above zero), followed by \textsc{spicker} and \textsc{durandal} (45/56 and 41/56
respectively).

We recorded the execution time of each algorithm: \textsc{durandal} was the
fastest (spending 140 minutes on the clustering of the whole dataset), then
Clusco on one CPU (426 min), \textsc{spicker} (435 min) and \textsc{calibur}
(856 min). If we allow for the possibility of parallel execution on GPU/CPU --
Clusco finished calculations in 131 min on 4 CPU’s, in 106 min on 4 CPU's and 1
GPU and in 47 min on 23 CPU’s. We summarized these results in Table 1.

Analyzing the above results, we can conclude that Clusco gives results which are
as good as the ones provided by the state-of-the-art \textsc{calibur} in half of
the time, however, on the commonly used today multicore machines, our program
gives results in the time about $18\times$ shorter than \textsc{calibur}.

\subsection*{Comparison of structures with reference/experimental model}

To compute the score between multi-model pdb file (tra.pdb) and the reference
structure (ref.pdb), one should run Clusco in the following way:

\texttt{clusco -t tra.pdb -e ref.pdb -s tmscore -o output.txt}

This command will compute the tm-score for each of tra.pdb models, saving the
results into output.txt. If \texttt{OMP\_NUM\_THREADS} variable was not set,
program will utilize all available CPU's.

We recorded the computation time of the tm-score to the reference (experimental)
structure with Clusco and the original TM-Score software\cite{Zhang2004a} using
the decoy set mentioned in the previous paragraph. TM-Score performed the task 
in 68 minutes, Clusco on 1 CPU -- in 53 minutes (speedup of about $1.2\times$),
but when we ran Clusco on 12 CPU’s, it completed the task in 7 minutes (speedup
about $10\times$) (detailed data in Additional file 1 Table 4).

It must be noted  that the computation time for GDT\_TS and MaxSub will be
mostly identical, since all of these algorithms use the same method for
selecting fragments of structure. Optionally, it is possible to compute more
exact GDT\_TS score with Clusco by using \texttt{ -s gdtExt} flag -- in this
particular case Clusco will split structures into many  more fragments.

For the comparison of cRMSD computation time, we used the \textsc{qcprot}
algorithm\cite{Theobald2005} claimed by the authors to be probably the fastest
available today. Recorded times  were only for the cRMSD routine  (without I/O
time). In this comparison test, we got slightly better results than the
\textsc{qcprot}: the speedup of $1.1-1.2\times$ for Clusco on one CPU and the
speedup of $12.7-16.1\times$ on one GPU. See Figure 1 for details.

Recently Hung \& Samudrala \cite{Hung2012} published an algorithm for the
computation of all-versus-all tm-score on \textsc{amd} GPU and CPU. We compared
Clusco with this algorithm using the exemplary data attached to the program
package (1000 models of 140 residues). Clusco on one CPU completed the
computation in 53.65 minutes, Hung \& Samudrala code -- in 57.18 minutes, but
Clusco can achieve pronounced speedup if executed in multi-CPU fashion (13.66
minutes on 4 CPU’s), which was not implemented in the Hung \& Samudrala
algorithm (see Figure 1 in Additional file 1 for tm-score values
comparison). However, users with access to the \textsc{amd} GPU can complete
this task significantly faster with Hung \& Samudrala algorithm.

\section*{Conclusion}

We presented here versatile software for comparison and clustering of protein
structures, optimized for novel multicore computers. We showed \textsc{cuda}
implementation of cRMSD algorithm which may be usable for creating of proteins
similarity matrices (a  bottleneck of the clustering software) as an input for
more efficient clustering algorithms. In the near future we will try to
implement the rest of score functions on the GPU.

Our software results in great-to-moderate speedup over an existing serial
execution algorithms, together with clustering results as good as obtained using
the state-of-the-art method, \textsc{calibur}.

Clusco is able to cluster small-to-moderate protein decoys with scoring
functions other than the cRMSD, i.e. the TM-Score, dRMSD, GDT\_TS, MaxSub,
Contact Map Overlap, especially on many-core machines, which is unique.

Clusco  may  be useful for protein modeling community as an all-in-one, fast and
easy in use software for daily lab work. It may be used as  a standalone program
for comparison or clustering of protein models or as a preprocessing tool for
clustering algorithms, either as a compiled program or a fragment of Clusco's
source code.

\section*{Competing interests}

The authors declare that they have no competing interests.

\section*{Availability and requirements}

 \textbf{Project name}:  ClusCo \\
 \textbf{Project home page}:  \url{http://biocomp.chem.uw.edu.pl/clusco} \\
 \textbf{Operating system}:  \textsc{gnu/Linux} \\
 \textbf{Programming language}:  \textsc{c/c++}, \textsc{cuda} \\
 \textbf{Other requirements}:  OpenMP library (included in \textsc{gcc}
$\geqslant 4.2$ compiler), optionally: \textsc{cuda sdk} and \textsc{cuda}
compatible graphic card \\
 \textbf{License}:  \textsc{gnu gpl} (scoring functions),
 Python License (Hierarchical Clustering library),
  custom license for K-means library (included in package) \\

\section*{Authors contributions}
MJ wrote algorithms, manual and manuscript. AK read, corrected and approved the
final manuscript.

\section*{Acknowledgements}
  \ifthenelse{\boolean{publ}}{\small}{}
We would like to thank Dr Sebastian Kmiecik for reading the manuscript.

MJ acknowledge the support from a Project operated within the Foundation for
Polish Science MPD Programme, co-financed by the EU European Regional
Development Fund. 

AK acknowledge support from the Foundation for Polish Science TEAM project
(TEAM/2011-7/6) co-financed by the European Regional Development Fund operated
within the Innovative Economy Operational Program.

%

{\ifthenelse{\boolean{publ}}{\footnotesize}{\small}
 \bibliographystyle{bmc_article}  
  \bibliography{cluscoref} }     

\ifthenelse{\boolean{publ}}{\end{multicols}}{}


\section*{Figures}
  \subsection*{Figure 1 - Comparison of running time of all-versus-all Clusco
and \textsc{qcprot}}
     cRMSD computation for three proteins of different length (71, 215 and 887
residues). For $N$ models it compute $N(N-1)/2$ cRMSD values.

\begin{center}
\includegraphics{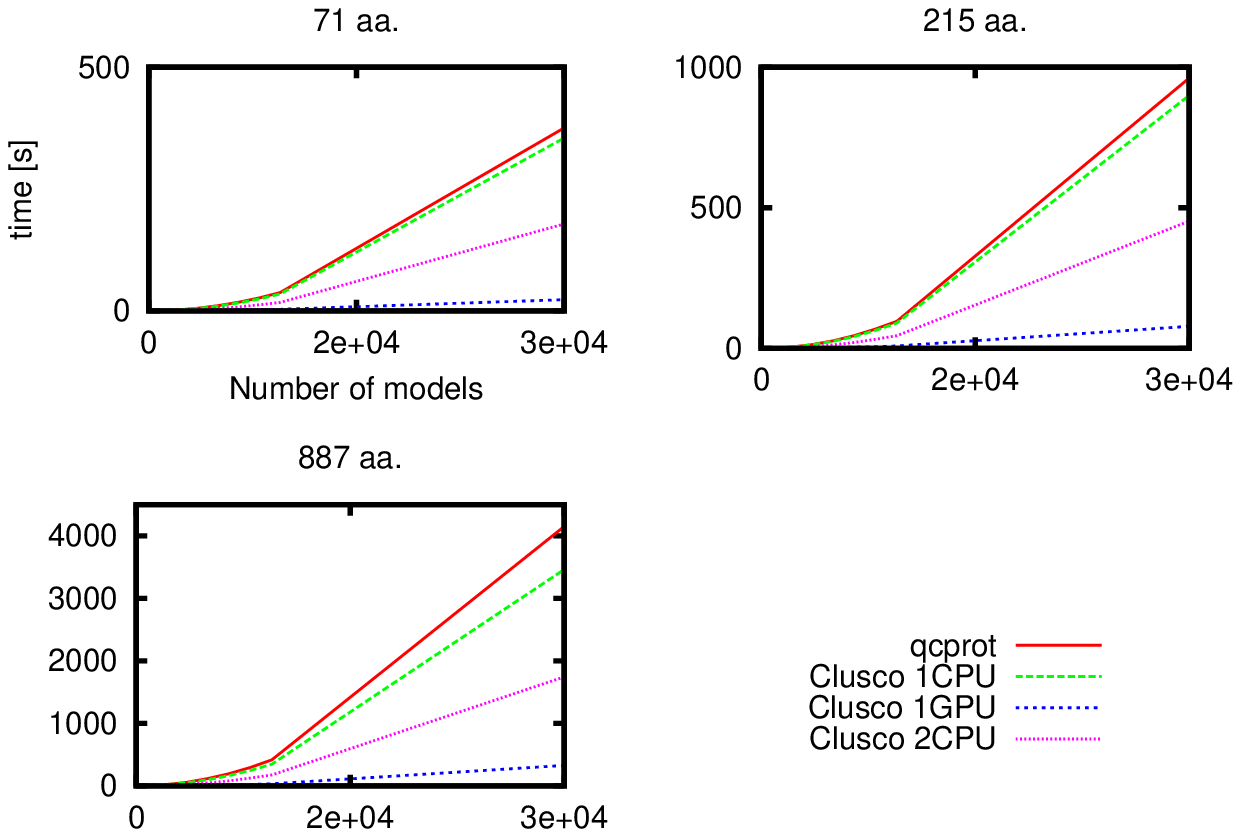}
\end{center}


\section*{Tables}
  \subsection*{Table 1 - Total time for clustering of decoys}
    \label{tbl-clust_time}  1thx\_ -- 32000 decoys of 108 aa. protein, 2reb\_2
-- 12500 decoys of 60 aa. protein. Need to note that \textsc{spicker} use
maximum of 13500 decoys. \par \mbox{}
    \par
    \mbox{\begin{tabular}{lrrr}
Program & Total time (min) & 1thx\_ & 2reb\_2  \\ \hline
\textsc{spicker} & 435 & 10  & 4 \\
\textsc{durandal} & 140 & 9 & 0.9\\
\textsc{calibur} & 859 & 64 & 1.2\\
Clusco 1CPU & 426 & 32 & 1.8 \\
Clusco 1CPU 1GPU & 213 & 19 & 0.7\\
Clusco 2CPU      & 219   &  16  &   0.9 \\
Clusco 2CPU 1GPU & 146    &  12  & 0.4   \\
Clusco 4CPU & 131 & 11 & 0.5 \\
Clusco 4CPU 1GPU & 106 & 7 & 0.4 \\
Clusco 23CPU & 47 & 3 & 0.3 \\ \hline
\end{tabular}}


\section*{Additional Files}
  \subsection*{Additional file 1 --- The Supporting Information}
    Additional comparison results with other software.

\end{bmcformat}
\end{document}